\documentstyle[12pt]{article}
\begin{document}
\title{Collective oscillations in superconductors revisited}
\author{S.N. Artemenko and A.F. Volkov}
\maketitle
\begin{abstract}
In the recent paper \cite{OT} Ohashi and Takada (OT) made statements that in
the clean limit considered by us \cite{AV}, weakly damped collective
oscillations in superconductors do not exist due to the Landau damping and
their spectrum differs from that obtained in Ref. \cite{AV}. In this Comment
we would like to note that these statements arise as a result of a
misunderstanding of the term "clean" case. OT considered the limit $\omega
\tau > 1$, meanwhile Artemenko and Volkov analysed the case $\tau T > 1$,
but $\omega \tau < 1$ (!). All these problems were discussed in the review
article \cite{UFN} which was, presumably, unknown to OT.  
\end{abstract}
Collective oscillations (CO) in the superconductors, the search for which had
continued since the fifties, were observed in experiments by Carlson and
Goldman \cite{Carl} more than two decades ago. The theoretical explanation
for weakly damped CO propagating in superconductors has been suggested in
Refs.\cite{AV},\cite{Sch}. Schmid and Sch\"{o}n \cite{Sch} considered the dirty
limit ($\tau T < 1$, $\tau$ is the elastic scattering time), and Artemenko
and Volkov \cite {AV} analysed the "clean" case ($\tau T > 1$). It was shown
that in both cases CO has a sound-like spectrum and
can exist only near $T_c$. The general theory of CO for both clean and dirty
cases was developed by Ovchinnikov \cite{Ovch} later. The oscillation
spectrum can be found from an equation analogous to the continuity equation
for, e.g. superfluid component and from an expression for the condensate
current. Since the condensate density has a different form in the dirty and
clean case, the velocity of CO and the range of their existence are
different. All these problems were discussed in detail in the review
articles \cite{UFN},\cite{KG},\cite{S}.
In spite of this there is still a misunderstnding in the literature about
CO. For example, in the recent paper \cite{OT} Ohashi and Takada
reconsidered the problem of CO (in their view the problem of CO in
superconductors is not solved yet) and made the following statements
1. CO in clean superconductors are strongly damped due to the Landau
damping,
2. the spectrum of the CO in clean superconductors differs from that
obtained in Ref. \cite{AV}.
In this Comment we would like to note that statement 1. is not new. The
importance of the Landau damping and the absence of the weakly damped CO in
the limit $\omega \tau > 1$ was noted in Ref. \cite {UFN}. Correspondingly,
the spectrum found in Ref. \cite {OT} in this limit does not coincide with
the spectrum found in Ref.\cite {AV} because the form of the spectrum in
this high frequency limit was not presented in Ref. \cite {AV} at all (it is
of little physical interest as mode is higly damped due to a strong
Landau damping). Therefore statement 2. arose due to misunderstanding of the
meaning "clean" case.  In Ref.\cite{AV} "clean" case meant that $\tau T >
1$, but $\omega \tau < 1$ (!). Meanwhile considering "clean" case Ogashi and
Takada mean the high fregency limit $\omega \tau > 1$.
Below we discuss briefly the problem of CO in superconductors and how their
spectrum depend on the screening by quasiparticles.
If the plasma frequency is smaller, than the energy gap $\Delta$ (like in
layered superconductors), the plasma mode continuosly transforms into the
Carlson-Goldman mode as temperature increases \cite{AK}. Plasma mode takes
place when frequency of dielectric relaxation due to quasiparticle currents
is smaller, than the plasma frequency, that is the perturbations of charge
densitiy are not screened. Then the superconducting current is compensated
by the displacemnt current. At higher temperatures the density of
quasiparticles becomes large enough to screen the charge density
oscillations totally, and superconducting current is compensated by the
current of quasiparticles. In this case oscillations of quasiparticle branch
imbalance play an important role. The latter case corresponds to the
Carlson-Goldman mode. However, it was shown \cite{AK97} that in the case of
d-pairing intensive relaxation of the branch imbalance due to elastic
scattering results in the strong damping of the Carlson-Goldman mode.
The statements made above may be illustrated by a transparent and very
simple way. Let us consider for simplisity a superconductor with s-pairing.
The expression for current density at $\omega \ll 1/\tau$ in the limit $Dk^2
\ll \omega$ has the form
\begin{equation}
j =\frac{c^2}{4\pi \lambda^2} P_s + \sigma_0\frac{\partial
P_s}{\partial t} -\sigma_1 \frac{\partial \mu}{\partial x} + \frac{1}{4\pi}
\frac{\partial E}{\partial t},
\label{j}
\end{equation}
where the first and the last terms describes the superconducting and the
displacement current, respectively. The second and the third terms describe
the quasiparticle current. Note that the response to the time derivative of
the superconducting momentum, {\em i. e.} of the gauge-invariant vector
potential ${\bf P}_s=(1/2)\nabla \chi - (1/c){\bf A}$, and to the gradient
of the gauge-invariant scalar potential $\mu =(1/2) (\partial \chi /\partial
t)+\Phi$ are described by different generalized conductivities. Thus from
the expression for the electric field ${\bf E_n} = - \nabla \mu_n-\partial
{\bf P}_s /\partial t$ one can see that the quasiparticle current can not be
discribed by the simple relation $j_{qp} = \sigma E$. At low temperatures,
$T \ll \Delta$, in the case of isotropic pairing these conductivities are
exponentially small, while near $T_c$, when $T \gg \Delta$ one gets
$\sigma_1 \approx \sigma_N$, $\sigma_0 \approx \sigma_N(1 + \Delta/T J)$, $J
\propto \ln{(\Delta/\omega)}$. Note that though the difference between
$\sigma_0$ and $\sigma_1$ near $T_c$ is small, it must not be neglected for
it determines the upper frequency limit of the range of low damping of the
CO.
To make the equations complete we need the equation to calculate potential
$\mu$ related to branch imbalance. Such an equation can be found from the
expression for the charge density \cite{UFN}, which can be presented in
physically transparent form 
\begin{equation} \frac{\partial \rho}{\partial
t} =\gamma(\frac{\partial }{\partial t} + \frac{1}{\tau_e})
\frac{\kappa^2}{4\pi}\mu + \sigma_1\frac{\partial^2
P_s}{\partial t \partial x} -\sigma_2 \frac{\partial^2 \mu}{\partial x^2}.
\label{rho}
\end{equation}
Here $\tau_e$ is the energy relaxation time. According to this equation
variations of charge density are created by variations of the branch
imbalance, {\em i. e.} of the difference between densities of electron-like
and of hole-like quasiparticles (the first term in the right-hand side), and
by the spatial variations of the quasiparticle flows. Again, the
quasiparticle flows are discribed by different "conductivities". At low
temperatures $\gamma \approx 1$, "conductivities" being exponentially small.
Near $T_c$ $\gamma =\pi \Delta/4 \pi$, $\sigma_2 \approx \sigma_N$.
The spectrum of eigenmodes can be found equating current $j$ to zero and
inserting charge density $\rho$ to the Poisson's equation. The character of
the spectrum depends on the relation between plasma frequency $\omega_p =
c/(\lambda \sqrt{\varepsilon}$ and frequency of dielectric relaxation
$\omega_r =4\pi \sigma/\varepsilon$.
At low temperatures $\omega_p \gg \omega_r$ quasiparticle conductivities are
small. From Poisson's equation we find $\mu = 0$, and from (\ref{j}) we get
$\omega^2 \approx \omega_p^2-\omega \omega_r$.  This equation is strict
provided the plasma frequency is smaller, than the gap, however, the result
is very similar in the opposite case as well.
At high temperatures, near $T_c$, $\omega_p \ll \omega_r$, and the
displacement current can be neglected and Poisson's equation reduces to the
neutrality condidtion. Thus, quasiparticles screen the perturbations of the
charge density, and quasiparticle current compensates the superconducting
current. This makes the mode soft. We find that the region of low damping is
limited by conditions $$\tau_e, \frac{c^2}{\sigma_N \lambda^2} \ll \omega
\ll \frac{c^2}{\sigma_N \lambda^2} \frac{T}{\Delta J}.$$
The spectrum in the range of small damping is given by $$\omega = Vq,\;
V^2=\frac{4\pi \sigma_N c^2}{\kappa^2 \lambda^2}.$$ Using in these equations
expressions for $1/\lambda^2$, describing the density of superconducting
electrons in the clean and dirty limits, $T_c\tau \gg 1$ and $T_c\tau \gg
1$, respectively, we get the results of papers \cite{AV} and \cite{Sch} for
corresponding limits: $$V =v\sqrt{\frac{7\zeta(3)\Delta}{3\pi^2T}}\;\;
\mbox{at}\;\; \tau_e, \frac{1}{\tau}\frac{\Delta}{T}^2 \ll \omega \ll
\frac{1}{\tau}\frac{\Delta}{T},$$ $$V =v\sqrt{\frac{2\tau\Delta}{3}}\;\;
\mbox{at} \;\; \tau_e, \frac{\Delta^2}{T} \ll \omega \ll  \Delta$$.
Note the analogy to the spectrum of phase oscillations in quasi
one-dimensional conductors with Charge-Density Wave. Similar to the case of
superconductors the Coulomb interactions hardens the Goldstone modes at low
temperatures \cite{LRA}, while at high temperatures screening by
quasiparticles softens the phason's spectrum making it sound-like both in
the case when quasiparticle scattering is neglected \cite{Kur}, and in the
collision dominated case \cite{AVp}. For detailed study of the phason
spectrum see \cite{AW}.  

\end{document}